\documentclass[a4paper,11pt,dvips]{article}
\usepackage{graphicx}
\usepackage{amssymb}
\usepackage{latexsym}

\begin{document}
\begin{titlepage}
\setcounter{page}{1}
\renewcommand{\thefootnote}{\fnsymbol{footnote}}

\begin{flushright}
\end{flushright}

\vspace{5mm}
\begin{center}

 {\Large \bf Electromagnetic Excitations of Hall Systems \\ on Four Dimensional Space }

\vspace{0.5cm}

{\bf Mohammed Daoud}$^{a,b}${\footnote{\sf  m$_-$daoud@hotmail.com}},  {\bf Ahmed Jellal}$^{c,d}$\footnote{{\sf ajellal@ictp.it} and
{\sf ahjellal@kfu.edu.sa}}  and  {\bf Abdellah Oueld Guejdi}$^e$

\vspace{0.5cm}

$^a${\em  Max Planck Institute for Physics of Complex Systems,
N\"othnitzer Str. 38,\\
 D-01187 Dresden, Germany}\\

{$^{b}$\em Department of Physics,
Faculty of Sciences, Ibn Zohr University},\\
{\em PO Box 8106,  80006 Agadir,
Morocco}

{$^c$\em Physics Department, College of Science, King Faisal University,\\
PO Box 380, Alahsa 31982,
Saudi Arabia}

$^d${\em  Theoretical Physics Group, Faculty of Sciences,
 Choua\"ib Doukkali University,\\
PO Box 20, 24000 El Jadida, Morocco}\\ 

$^e${\em Department of Mathematics, Faculty of Sciences, University Ibn Zohr,\\
PO Box 8106, Agadir,
Morocco}\\[1em]

\vspace{3cm}

\begin{abstract}

The noncommutativity of a four-dimensional phase space is introduced from a purely
symplectic point of view. We show that there is always a coordinate map to locally
eliminate the gauge fluctuations inducing the deformation of the symplectic structure.
This uses the Moser's lemma; a refined version of the celebrated Darboux theorem.
We discuss the relation between the coordinates change arising from Moser's
lemma and the Seiberg--Witten map. As illustration, we consider the quantum
Hall systems on ${\bf CP}^2$. We derive the action describing  the electromagnetic
interaction of Hall droplets. In particular, we show that the velocities of the edge
field, along the droplet boundary, are noncommutativity parameters-dependents.

\end{abstract}
\end{center}
\end{titlepage}

\newpage
\section{Introduction}

Recently, there has been considerable interest in the noncommutative geometry as
framework for physical theories and as tool for study certain mathematical structures,
which appears in some physical models. This is mainly motivated by the new development in
string theory [1]. Subsequently, the idea of  non commutative space time at small length scales [2]
 has been drawn much
attention in various fields and found interesting implications,
see for instance [3-4].

Since the noncommutative space resembles a quantum phase space
(with noncommutativity parameter $\theta$ playing the role of $\hbar$),
many papers have been devoted to study
various aspects of quantum mechanics [5-9] on the noncommutative space
where space-space is non commuting and/or momentum-momentum is non commuting.
The usual way of investigating the noncommutative quantum mechanics is to map the noncommutative space
to a commutative one. At classical level, this map turns out to be similar to
the celebrated Darboux transformation. In this respect, the noncommutative quantum mechanics
can be viewed as quantization of a phase space equipped with modified symplectic structure.
To eliminate the fluctuation, one has to define a diffeomorphism, which maps the modified symplectic form
to its counter part in the commutative case. Hence, one of the main aims of the present work is to give
a general prescription to perform this "dressing" transformation for arbitrary modified closed two-form
on a curved phase space. This prescription uses
the Moser's lemma [10] which is a refined version of Darboux theorem.
We will discuss many facets and consequences of this transformation.
We also compare this method with the transformation, which arises from the Hilbert--Shmidt
orthonormalization method in four-dimensional phase space.

 On the other hand, the prototypical topic at the interface between the noncommutative geometry and
 condensed matter physics was in the last decade, the quantum Hall effect. Indeed,
according to the Laughlin [11], a large collection of fermions in a strong magnetic field behaves
like a rigid droplet of liquid. This  incompressible quantum
fluid picture  constitutes the
basis of the main advances in this field of research, especially
its connection with the noncommutative structures. Indeed, it was shown
that Laughlin states at filling
factor  $1/k$ can be provided by an appropriate noncommutative
finite Chern--Simons matrix model at level $k$ and hence reproduces
the basic features of quantum Hall states [12-13].
In connection with quantum Hall systems in higher dimensions [14-25],
the ideas of the noncommutative geometry were useful to show that the effective action for the edge
excitations of a quantum hall droplet is generically given by a
chiral boson action [21-25]. In relation with these issues, the second main task of this paper concerns
the electromagnetic excitations of Hall droplets in four-dimensional
complex projective space. The electromagnetic field is introduced
as a variation of the ${\bf CP}^2$ symplectic two-form.

The outline of the paper is as follows. In section 2, we first review the basic structure of quantum systems whose
elementary transitions (excitations) operators close the Lie algebra $su(d+1)$. We define the Bargmann phase space
and the corresponding symplectic  structure $\omega_0$
of such system. This is realized by making use of the coherent states formalism,
  which offers a  very nice way  in the study of the quantum classical
correspondence. We introduce the noncommutative Bargmann space by shifting the symplectic two-form
$\omega_0 \longrightarrow \omega_0 + F$ where $F$ is the perturbation induced by a external gauge field.
Consequently, the position as well as momenta coordinates cease  to Poisson commute.
Thus, to study the dynamics of a given system whose phase space is noncommutative, it is more appropriate
to find out a dressing transformation that converts the modified symplectic form to $\omega_0$.
This issue is presented in section 3. We give a general procedure based on the Moser's lemma to
eliminate the fluctuations of the symplectic structure. This generalizes the maps based on the Darboux transformations
to include also curved phase spaces. The effects of the modification become then encoded in the Hamiltonian of the system.
We discuss the relation between the obtained  transformation and the famous Seiberg--Witten map, which was initially
introduced in the context of the noncommutative gauge theory [1], see also [26-28].  In section 4, we treat the case
where the matrix elements of the fluctuation form $F$ are constants. We show that, in this particular case, one can obtain
an exact dressing transformation contrarily to Moser's procedure (which is in some sense perturbative).
This exact transformation is similar to Hilbert--Schmidt orthonormalization procedure. As illustration of our results,
we consider, in Section 5, the problem of the electromagnetic excitations of a quantum Hall droplet in the complex
projective space ${\bf CP}^2$. The coupling of the quantum Hall droplet with electromagnetic field is done from a
purely symplectic point of view. We give the Wess--Zumino--Witten action describing the edge excitations
on the boundary of the quantum Hall droplet. We show that the electromagnetic field modify
the velocities of the propagation of the chiral field along the angular directions. Concluding remarks close the present paper.

 \section{Symplectic deformation and noncommutative Bargmann space}

 \subsection{General considerations}

It is well established that for an exact solvable quantum system,
 there is always a well-defined group structure. We denote by ${\cal G}$
 the corresponding operator algebra. The dynamical properties of this system are described within a
Hilbert space ${\cal F}$ and the dynamical observables are represented
by operators acting on it. This space is completely
specified by determining the subset of ${\cal G}$ generated by the
elementary transition or excitation operators of the system, i.e. annihilation $t_i^-$
 and creation  $t_i^+$, with  $i = 1, 2, \cdots ,
d$.  The Hamiltonian system and  various
transition operators can be expressed in terms of the scale operators.

On the other hand, for a classical system, the dynamical observables
are differential analytic functions defined on a phase space endowed
with a symplectic structure. The classical limit can occur only if
such structure can emerges from the quantum system in question. In
other words, one must construct a geometry originated from the
Hilbert space, which   must possess the necessary
symplectic structure. Indeed, for a quantum system, namely an
algebraic structure $({\cal G}, {\cal F})$, there exist $2d$-dimensional
symplectic manifold ${\cal M}$, which is isomorphic to the
so-called coset space $G/H$, where $G$ is the covering group of
${\cal G}$ and $H$ is the maximal stability subgroup of $G$ with
respect to the fixed state $\vert \psi_0 \rangle$, i.e. the highest
weight vector.

In the present analysis, we mainly focus on the $su(d+1)$ quantum systems.
For the Lie algebra $su(d+1)$, there are $2d$  generators,  which are not
in its subalgebra $u(d)$. These can be separated into the lowering $t_{-i}$ and
 raising $t_{+i}$ types. It is interesting to note
that  $su(d+1)$ can be introduced through the Weyl generators $t_{\pm i}$ and the
triple commutation relations, such as
\begin{eqnarray}
&&[[t_{+i} , t_{-j} ], t_{+k} ] = \delta_{jk}t_{+i} +\delta_{ij}t_{+k}\\
&&[[t_{+i} , t_{-j} ], t_{-k} ] = -\delta_{ik}t_{-j} - \delta_{ij}t_{-k}
\end{eqnarray}
implemented by the mutual commutators
\begin{equation}
[t_{+i} , t_{+j} ] = 0, \qquad [t_{-i} , t_{-j} ] = 0.
\end{equation}
 Recall that, the mentioned description 
was introduced for the first time by Jacobson [29] in the context of Lie triple systems. This provides a minimal alternative
to the Chevally description. The corresponding Hilbert space [30], see also [31-33], is
\begin{equation}
{\cal F} = \left\{ \vert n_1, n_2, \cdots , n_d\rangle; \ \ n_i \in {\mathbb N}\right\}.
\end{equation}
 The elementary excitations operators
act on ${\cal F}$ as
\begin{equation}
t_{\pm i} \vert n_1, \cdots, n_i, \cdots, n_d\rangle\ = \sqrt{F_i(n_1, \cdots, n_i \pm 1, \cdots , n_d)} \vert n_1, \cdots , n_i\pm 1 , \cdots  n_d\rangle\
\end{equation}
where the structure function $F(n_1,\cdots , n_i, \cdots , n_d)$ is given by
\begin{equation}
 F_i(n_1,\cdots , n_i, \cdots , n_d) = n_i\left[ k + 1 - (n_1+n_2+\cdots n_d)\right]
\end{equation}
and $k$ is a real number labeling the representation.  The Hilbert space has a finite dimension if
the quantum numbers $n_i$ fulfilled the condition
$  (n_1+n_2+\cdots n_d) \leq k$. This dimension is
$$ {\rm dim} ~ {\cal F} = \frac{(k+d)!}{k!d!}$$
which is nothing but the dimension
of the symmetric representations of the Lie algebra $su(d+1)$.

To obtain
the manifold ${\cal M}$, one can use an unitary
exponential mapping. This is
\begin{equation}
\sum_{i=1}^{d}(\eta_i t_{+i} - \bar \eta_i t_{-i}) \longrightarrow
\Omega = \exp \sum_{i=1}^{d}(\eta_i t_{+i} - \bar \eta_i t_{-i})
\end{equation}
where $\eta_i$ are complex parameters and $\Omega$ is an unitary coset
representative of the coset space $G/H \equiv SU(d+1)/U(d)$. This gives the complex projective space ${\bf CP}^d$ as geometrical realization corresponding to
${\cal F}$.  This correspondence can be better visualized using the formalism of generalized coherent states of $G$, such as
\begin{equation}
\Omega \longrightarrow \vert \Omega \rangle \equiv \Omega \vert \psi_0 \rangle = \Omega \vert
0, 0, \cdots , 0 \rangle.
\end{equation}
This gives (see for instance in [33] where the notations are more or less similar)
\begin{equation}
 \vert \Omega \rangle = \sum_{\{ n_i \}} \bigg[\frac{k!}{n_1n_2!\cdots n_d!(k - n)!}\bigg]^{\frac{1}{2}}
\frac{z^{n_1}_1 z^{n_2}_2 \cdots z^{n_d}_d}{(1+\bar z\cdot z)^{k/2}}\vert n_1, n_2, \cdots, n_d\rangle\
\end{equation}
where $n = n_1+ n_2 +\cdots +n_d$ and the complex variables are $ z_i = \frac{\eta_i}{\sqrt{\bar \eta . \eta}}\tan \sqrt{\bar \eta \cdot \eta}$.
Obviously, these states constitute an complete set with respect to the measure 
 \begin{equation}
 d\mu ( \bar z , z) = \frac{(k+d)!}{\pi^d k!} ~ \frac{d^2z_1d^2z_2 \cdots d^2z_d}{(1+\bar z\cdot z)^{d+1}}.
\end{equation}
The space of analytical functions (Bargmann space) defined by the above coherent states is equipped with
a symplectic (Khaler) two-form. This makes it into classical phase space and hence it connects the quantum model
to its semiclassical limit. It can be realized by introducing the Kahler
potential
 \begin{equation}
K_0( \bar z , z) =  \ln \vert \langle \psi_0 \vert \Omega \rangle \vert^{-2}  = k \ {\ln}( 1 + \bar z\cdot z)
\end{equation}
which allows us to define a closed symplectic two-form
\begin{equation}
 \omega_0 = i g_{i\bar j} dz^i\wedge d\bar z^j.
\end{equation}
The corresponding Poisson bracket is given by
\begin{eqnarray}
\{ f , g \} = -i g^{i\bar j} \left(\frac{\partial f}{\partial
z^i}\frac{\partial g}{\partial \bar z^j} - \frac{\partial
g}{\partial z^i}\frac{\partial f}{\partial \bar z^j}\right).
\end{eqnarray}
The components of the metric tensor take the form
$$ g_{i \bar j} = \frac{\partial^2 K_0(\bar z , z )}{\partial z_i \partial \bar z_j} =
k (1 + \bar z\cdot z)^{-2}[(1 + \bar z\cdot z)\delta_{ij} - \bar z_iz_j]$$
and therefore the matrix elements of its inverse are
$$ g^{i\bar j} = \frac{1}{k} (1 + \bar z\cdot z) (\delta_{ij} + z_i\bar z_j).$$
By introducing the canonical coordinates $(q,p)$ of $G/H = SU(d+1)/U(d)$
\begin{eqnarray}
\frac{1}{\sqrt{2 k}}(q_i + i p_i) = \frac{z_i}{\sqrt{1 + \bar z\cdot z}}
\end{eqnarray}
it is easily seen that the Poisson
two-form can be transformed into the canonical one. This is
\begin{eqnarray}
\omega_0 = \sum_i dq_i\wedge dp_i.
\end{eqnarray}
Now the Poisson bracket becomes
\begin{eqnarray}
\{ f , g \} =  \sum_{i=1,2}\left(\frac{\partial f}{\partial
q^i}\frac{\partial g}{\partial p^i} - \frac{\partial g}{\partial
p^i}\frac{\partial f}{\partial  q^i}\right)
\end{eqnarray}
This re-parametrization offers a familiar phase space structure with $ \sum_i (p_i^2 + q_i^2) \leq 2k$,
 which shows that
the phase space of the system is compact.  As mentioned in the introduction,
we will essentially interested by the four-dimensional phase space, namely $d=2$ in the above analysis.

\subsection{{Deformed symplectic structure }}

We now assume that the symplectic structure of the phase space is
modified due to the presence  of an external electromagnetic
background. This can be formulated by replacing the canonical
two-form $\omega_0$ by a closed new one, such as
\begin{eqnarray}
\omega = \omega_0 + F = \omega_0 - \frac{1}{2} {\cal B}_{ij}(q)
dq^i\wedge dq^j + \frac{1}{2}{\cal E}_{ij}(p) dp_i\wedge dp_j
\end{eqnarray}
where the deformation is encoded in the  antisymmetric
tensors ${\cal E}_{ij}$ and ${\cal B}_{ij}$. This modification requires
a condition on the space dimension, namely $ d > 1$.
Note that, $\omega$
can be mapped, in a compact form, as
\begin{eqnarray}
\omega = \frac{1}{2}   \omega_{IJ}(\xi) d\xi^{I} \wedge d\xi^{J}
\end{eqnarray}
where $I, J = 1, 2, 3, 4$, with $\xi^i = q^i$ and $\xi^{i+2} = p^i$
for $i = 1,2$. The nonvanishing elements of the antisymmetric
matrix $\omega$ are
\begin{eqnarray}
\omega_{12} = - {\cal B}_{12}, \qquad \omega_{34}=  {\cal
E}_{12}, \qquad \omega_{13}= \omega_{24} = 1.
\end{eqnarray}
It is nondegenerate\ i.e. ${\det}~\omega \neq 0$, when the antisymmetric
tensors ${\cal E}_{ij}$ and ${\cal B}_{ij}$ satisfy the condition
$\det( 1_{2\times 2} - {\cal E}{\cal B}) \neq 0$. This conclusion can  easily
be reached by writing $\omega$ in terms of matrix. Here we assume
that such a condition is satisfied. To find the classical equations
of motion and establish the connection between the classical and
quantum theory, it is necessary to define the Poisson brackets
associated with the new phase space geometry in a consistent way.
Indeed, since the Poisson brackets for the coordinates on
the phase space are the inverse of the symplectic form as matrix, we
have
\begin{eqnarray}
\{{\cal F} , {\cal G}\} =  (\omega^{-1})^{IJ}\frac{\partial {\cal
F}}{\partial \xi^{I}}\frac{\partial {\cal G}}{\partial \xi^{J}}
\end{eqnarray}
where $(\omega^{-1})^{IJ}$ is the inverse matrix of $\omega_{IJ}$
(17) and $({\cal F},{\cal G})$ are two functions defined on the phase
space.  After a straightforward calculation, one can show
\begin{eqnarray}
\{{\cal F}, {\cal G}\} = \sum_{ik}
(\Theta^{-1}_1)_{ik}\frac{\partial {\cal F}}{\partial q^i}\left[
\frac{\partial {\cal G}}{\partial p^k} - \sum_{j} {\cal
E}_{kj}\frac{\partial {\cal G}}{\partial q^j} \right] -
(\Theta^{-1}_2)_{ik} \frac{\partial {\cal F}}{\partial p^i}\left[
\frac{\partial {\cal G}}{\partial q^k}- \sum_j{\cal
B}_{kj}\frac{\partial {\cal G}}{\partial p^j}\right]
\end{eqnarray}
where the matrix elements of $\Theta_1$ and $\Theta_2$  are defined
by
\begin{eqnarray}
&&(\Theta_1)_{ij} = \delta_{ij} - {\cal E}_{ik}{\cal B}_{kj}\\
&& (\Theta_2)_{ij} = \delta_{ij} - {\cal B}_{ik}{\cal E}_{kj}.
\end{eqnarray}
They can also be read in matrices form as $\Theta_1 = 1 - {\cal
E}{\cal B}$ and $\Theta_2 = 1 - {\cal B}{\cal E}$, respectively. It
follows that, the modified canonical Poisson brackets are
\begin{eqnarray}
&& \left\{ q^i , q^j \right\} = - \sum_k(\Theta^{-1}_1)_{ik}{\cal E}_{kj}\\
&& \left\{ p^i , p^j \right\} =  \sum_k(\Theta^{-1}_2)_{ik}{\cal B}_{kj}\\
&& \left\{ q^i , p^j \right\} =  (\Theta^{-1}_1)_{ij} = (\Theta^{-1}_2)_{ji}.
\end{eqnarray}
These relations traduce the noncommutativity of the phase space generated by the symplicric modification.
Clearly, in the limiting case ${\cal E} = 0$ and ${\cal B} = 0$,
the noncommutative relations (24-26) reduce to the canonical Poisson brackets.
According to  the modified  symplectic structure of the
phase space, we introduce the vector fields $X_{\cal F}$ associated
to a given function  ${\cal F}(q^i , p^j)$. This is
\begin{eqnarray}
X_{\cal F} = \sum_{i} X^i \frac{\partial }{\partial q^i} +
Y^i\frac{\partial }{\partial p^i}
\end{eqnarray}
such that the interior contraction of $\omega$ with  $X_{\cal F}$
gives
\begin{eqnarray}
{\iota }_{X_{\cal F}}\ \omega = d {\cal F}.
\end{eqnarray}
A simple calculation leads
\begin{eqnarray}
 && X^i= \sum_j(\Theta^{-1}_1)_{ij}\left(\frac{\partial {\cal
F}}{\partial p^j} - \sum_k{\cal E}_{jk} \frac{\partial {\cal F}
}{\partial q^k}\right)\\
 && Y^i = - \sum_j(\Theta^{-1}_2)_{ij}\left(\frac{\partial {\cal
F}}{\partial q^j} - \sum_k{\cal B}_{jk} \frac{\partial {\cal F}
}{\partial p^k}\right).
\end{eqnarray}
One can check
\begin{eqnarray}
{\iota }_{X_{\cal F}}{\iota }_{X_{\cal G}} \omega = \{{\cal F}, {\cal
G}\}.
\end{eqnarray}

\section{Noncommutative dynamics in Bargmann space }

The celebrated Darboux theorem guarantees the existence of local
coordinates $(Q_i, P_i)$ such that $\omega$ takes a canonical form.
Such Darboux coordinates transformation are easily
obtained once of the tensors ${\cal B}$ and ${\cal E}$ vanishes.
This can be done by using one-form potential $A_i(q)dq_i$
 and $\bar A_i(p)dp_i$ that defines a $U(1)$ abelian
potential $A$. It is
\begin{eqnarray}
F = dA, \qquad A =  A_Id\xi^I = A_i(q) dq^i + \bar{A}_i(p) dp^i
\end{eqnarray}
 where bar is just a notation and has nothing to do with the usual complex
conjugate. Consequently, for ${\cal E} = 0$, the Darboux coordinates are given by
\begin{eqnarray}
Q_i = q_i, \qquad P_i = p_i - A_i(q).
\end{eqnarray}
However, for ${\cal B} = 0$, one obtains
\begin{eqnarray}
Q_i = q_i + \bar{A}_i(p) \qquad P_i = p_i .
\end{eqnarray}
In the case where both of forms ${\cal B}$ and ${\cal E}$ are constant, $\omega$ can
be re-written in canonical form.
This can be achieved by making use of
 a linear symplectic orthonormalization procedure \`a la Hilbert
Schmidt, which will be treated in section 4. However, for nonconstant ${\cal B}$ and ${\cal E}$, the Darboux procedure fails
in converting the symplectic two-form $\omega_0 + F$ in Darboux canonical form. As alternative
method,  one has to employ is based on the Moser's lemma, which constitutes a refined version of Darboux theorem.
This will be detailed in what follows.

\subsection{Symplectic dressing through Moser's lemma}

Let us start by revisiting the derivation of  Moser's lemma which behind a nice procedure to locally eliminate
the fluctuation ${\cal {E+B}}$ of the initial symplectic two form $\omega_0$.
To give a general algorithm to realize a dressing transformation through Moser's lemma,
we will consider the general case where the  matrix elements of  $\omega_0$  are phase space dependents.

 According to Moser's
lemma, there  always exists a diffeomorphism on the phase space $\phi$ whose pullback maps $\omega$ to $\omega_0$. This is
\begin{eqnarray}
\phi^{\ast}(\omega_0 + F) = \omega_0
\end{eqnarray}
namely, we have
\begin{eqnarray}
\phi  : \xi^I \longmapsto \phi(\xi^I), \qquad \frac{\partial\phi(\xi^K)}{\partial\xi^I}
\frac{\partial\phi(\xi^L)}{\partial\xi_J} \omega_{KL}(\phi(\xi))= {\omega_0}_{IJ}(\xi).
\end{eqnarray}
To find out this change of coordinates, one can start by defining a family of one parameter  of symplectic forms
\begin{eqnarray}
\omega(t) = \omega_0 + t F
\end{eqnarray}
interpolating $\omega_0$ and $\omega_0 + F$ for $t=0$ and $t=1$, respectively, with
 $0 \leq t \leq 1$. Note that, $t$ is just an affine parameter labeling the flow generated by a smooth $t$-dependent vector field $X(t)$.
Accordingly, one  also define a family
of diffeomorphisms
\begin{eqnarray}
\phi^{\ast}(t)\omega(t) = \omega_0
\end{eqnarray}
satisfying $\phi^{\ast}(t=0)= id$ and  $\phi^{\ast}(t=1)$ will  be
the required solution of our problem, i.e. (35).
 Differentiating (38), one check that $X(t)$  must satisfy the identity
\begin{eqnarray}
0 = \frac{d}{dt} \left[\phi^{\ast}(t)\omega(t)\right] =
\phi^{\ast}(t)\left[L_{X(t)}\omega(t) + \frac{d\omega(t)}{dt}\right].
\end{eqnarray}
where $L_{X(t)}$ denotes the Lie derivative of the field $X(t)$.
Using the Cartan identity $L_{X} = ~ \iota_X\circ d + d\circ \iota_X$
and the fact that $d\omega (t) = 0$, we obtain
\begin{eqnarray}
 \phi^{\star}(t)\left\{d\left[\iota_{X(t)}\omega(t)\right] + F\right\}=0
\end{eqnarray}
where $\iota_X$ stands for interior contraction as above. It
follows that  $X(t)$ is verifying the linear
equation
\begin{eqnarray}
\iota_{X(t)}\omega(t) + A = 0
\end{eqnarray}
which  solves (39). Therefore, the components of  $X(t)$ are given by
\begin{eqnarray}
X^I(t) = - A_J \omega^{-1 JI} (t).
\end{eqnarray}

 For small fluctuations of the symplectic structure, i.e. $F \ll \omega_0$, one can write the inverse of $\omega$ as
\begin{eqnarray}
 \omega^{-1}(t) = \omega_0^{-1} - t\omega_0^{-1} F \omega_0^{-1} + t^2 \omega_0^{-1} F \omega_0^{-1} F \omega_0^{-1} + \cdots.
\end{eqnarray}
This determines the components of  $X(t)$ in terms of the $U(1)$ connection $A$
and its derivatives and allows us to write down the explicit form of the transformation $\phi$. Indeed,
since the $t$ evolution of $\omega(t)$ is governed by the first order differential equation
\begin{eqnarray}
 \left[\partial _t + X(t)\right]\omega(t) = 0
\end{eqnarray}
 it is easy to show that
\begin{eqnarray}
 \left[\exp(\partial _t + X(t))\exp(-\partial _t )\right]\omega(t+1) = \omega(t).
\end{eqnarray}
This leads to the relation
\begin{eqnarray}
 [\exp(\partial _t + X(t))\exp(-\partial _t )]\vert_{(t=0)}(\omega_0 + F) = \phi^{\ast}( \omega_0 + F ) = \omega_0
\end{eqnarray}
where $\phi^{\ast}$ is given by
\begin{eqnarray}
\phi^{\ast} = id + X(0) + \frac{1}{2} (\partial_t X)(0) + \frac{1}{2} X^2(0) + \cdots.
\end{eqnarray}
More explicitly, using (42), the contribution arising from the second term in (47) read as
\begin{eqnarray}
X(0) = \omega_0^{-1 IJ}A_J\partial_I.
\end{eqnarray}
The contribution of the third term in (47) is
\begin{eqnarray}
\frac{1}{2} (\partial_t X)(0) = -\frac{1}{2}(\omega_0^{-1} F \omega_0^{-1})^{IJ}A_J\partial_I.
\end{eqnarray}
The last term in (47) gives
\begin{eqnarray}
\frac{1}{2} X^2(0) = \frac{1}{2}(\omega_0^{-1IJ}A_J\partial_I)(\omega_0^{-1I'J'}A_{J'}\partial_{I'})
\end{eqnarray}

Finally, in terms of local coordinates, the coordinate transformation $\phi$
whose pullback maps $\omega_0 + F \longrightarrow \omega_0$
is given by
\begin{eqnarray}
\phi (\xi^L) = \xi^L + \xi^L_1 + \xi^L_2 + \cdots
\end{eqnarray}
where $\xi^L_1$ is
\begin{eqnarray}
\xi^L_1 = \omega_0^{-1 LJ}A_J
\end{eqnarray}
and $\xi^L_2 $ takes the form
\begin{eqnarray}
\xi^L_2 = - \frac{1}{2}\omega_0^{-1 LK} F_{KL'}\omega_0^{-1 L'J}A_{J}
+\frac{1}{2}\omega_0^{- 1IJ}A_J(\partial_I\omega_0^{-1 LJ'})A_{J'} +
\frac{1}{2}\omega_0^{-1 IJ}A_J\omega_0^{-1 LJ'}(\partial_IA_{J'}).
\end{eqnarray}
Using the relations
\begin{eqnarray}
&& \partial_{J'}A_{I'} = (\partial_{J'}\omega_{0I'I})\xi^{I}_1 + \omega_{0I'I}(\partial_{J'}\xi^{I}_1)\\
&& \partial_I\omega_0^{-1 LJ'} = -\omega_0^{- 1 LJ"}(\partial_I\omega_{0 J"K})\omega_0^{- 1 KJ'}
\end{eqnarray}
and the antisymmetry property of the symplectic form, keep in mind  that $\omega_0$ is assumed closed and nonconstant,
one can check
\begin{eqnarray}
\xi^L_2 &=& -\omega_0^{-1 LK}F_{KL'}\xi^{L'}_1 + \frac{1}{2} \omega_0^{-1 LK}
\omega_0^{-1 MJ}A_J \omega_0^{-1 NJ'}A_{J'}\partial_M\omega_{0 NK} 
\nonumber\\
&& + \frac{1}{2}
\omega_0^{-1 LK} \omega_0^{-1 MS}A_S \omega_{0 MN}
\partial_K(\omega_0^{-1 NS'}A_{S'}).
\end{eqnarray}
It is remarkable that this dressing transformation  coincides with the
Susskind map derived in connection with  the quantum Hall systems and
noncommutative Chern--Simons theory [12]. It leads also to  the very familiar
Seiberg--Witten map [1] used in the context of the string and noncommutative gauge theories.
This will be clarified in the next subsection.

\subsection{{Seiberg--Witten map in four-dimensional phase space}}

In fact, one can see from (52) and (56) that the dressing transformation can
be written as
\begin{eqnarray}
\phi (\xi^L) = \xi^L +  {\hat A}^L
\end{eqnarray}
where we have set
\begin{eqnarray}
{\hat A}^L &=& { \omega_0}^{-1LK}\bigg[A_K
- F_{KL'}\omega_0^{-1 L'M}A_M + \frac{1}{2} {\omega_0}^{-1 MJ}A_J {\omega_0}^{-1 NJ'}A_{J'}\partial_M\omega_{0NK}
\nonumber\\
&& + \frac{1}{2}  {\omega_0}^{-1 MS}A_S \omega_{0 MN} \partial_K ({\omega_0}^{-1 NS'}A_{S'})\bigg].
\end{eqnarray}
The transformation (57) is similar to the so-called Susskind map. It encodes the
geometrical fluctuations induced by the external magnetic field $F$. Also,  it coincides  with
the Seiberg--Witten map
in a curved manifold  for the noncommutative abelian gauge theory [30]. Indeed, under the gauge transformation
\begin{eqnarray}
A \longrightarrow A + d\Lambda
\end{eqnarray}
 the components (58) transform as
 \begin{eqnarray}
{\hat A}^L \longrightarrow {\hat A}^L +
 \omega_0^{- 1 LJ}\partial_J {\hat \Lambda} + \{ {\hat A}^L , {\hat \Lambda}\}+ \cdots
\end{eqnarray}
where the noncommutative gauge parameter $\hat \Lambda $
\begin{eqnarray}
{\hat \Lambda} = \Lambda + \frac{1}{2} \omega_0^{-1 IJ}A_J\partial_I\Lambda + \cdots
\end{eqnarray}
is written as function of  $\Lambda$ and the abelian connection $A$. The equations (58), (60) and (61)
are the semiclassical versions of the Seiberg--Witten map.
The connection $\hat A$ is the induced noncommutative gauge potential given in
terms of its commutative counter part $A$. This establish a  correspondence between
symplectic deformations and non commutative gauge theories.

Now we return to the situation of our purpose where the phase space is
four-dimensional and equipped with the canonical Darboux form $\omega_0$ given in (15).
In this particular case, one can verify, by using
(32), (51), (52) and (56), that   the deformed two-form $\omega_0 + F$ (17) takes the canonical
form
\begin{eqnarray}
\omega_0 + F =   dQ^i \wedge dP^i
\end{eqnarray}
where the new phase space variables $Q^i$ and $P^i$ are given by
\begin{eqnarray}
&&Q^i = \phi^{-1}(q^i) = q^i + \bar A_i(p) - \sum_{j=1,2}A_j(q)\left[
{\cal E}_{ij}(p) - \frac{1}{2}\frac{\partial \bar A_j(p)}{\partial
p_i}\right]+ \cdots
\\
&&P^i = \phi^{-1}(p^i) =  p^i - A_i(q) + \sum_{j=1,2}\bar A_j(p)\left[
{\cal B}_{ij}(q) + \frac{1}{2}\frac{\partial  A_j(q)}{\partial
q_i}\right]+ \cdots.
\end{eqnarray}
It is interesting to note that for  $\bar A_i(p) = 0$ (respectively $A_i(q) = 0$) we obtain (33)
(respectively (34)) and recover the  Darboux transformations discussed above when one of the tensors ${\cal B}$ and ${\cal E}$ vanishes.
On the other hand, when the gauge potential (32) is defined as
\begin{eqnarray}
A = -\frac{1}{2} \left(\bar \theta \epsilon_{ij}q_idq_j - \theta
\epsilon_{ij}p_idp_j\right)
\end{eqnarray}
corresponding to a constant electromagnetic fields $F$ ($\theta$ and $\bar \theta$ real constants), the dresssing transformation  (63-64) gives
\begin{eqnarray}
&& Q^i = \left(1 + \frac{3}{8} \theta \bar\theta\right) q^i +
\frac{\theta}{2}\sum_k\epsilon_{ki}p^k\\
&&  P^i = \left(1 + \frac{3}{8} \theta \bar\theta\right) p^i +
\frac{\bar\theta}{2}\sum_k\epsilon_{ki}q^k.
\end{eqnarray}
$\epsilon_{ij}$, appearing in (65), is the usual antisymmetric tensor,
namely $\epsilon_{12} = - \epsilon_{21} = 1$.

\subsection{ Hamiltonian system}

Let ${\cal H}\equiv {\cal H}(p,q)$ to be the original classical Hamiltonian.
In modifying the symplectic structure, the dynamics becomes
described by two-form $\omega_0 + F$. The dressing transformation converts the
 dynamical system of $(\omega_0 + F, {\cal H})\Big|_{qp}$ to $(\omega_0 , {\cal H}_A)\Big|_{QP}$ where
we use the old symplectic form but  a different Hamiltonian,
 which can be obtained by simply replacing the old phase space
 variables in terms of the new ones. In this respect, using (57) (or inverting (63) and (64)), one obtains
\begin{eqnarray}
&& q^i = \phi (Q^i) = Q^i - \bar A_i(P) + \sum_{j=1,2}A_j(Q)\bigg[
{\cal E}_{ij}(P) - \frac{1}{2}\frac{\partial \bar A_j(P)}{\partial
P_i}\bigg]+ \cdots\\
&&p^i = \phi (P^i) =  P^i + A_i(Q) - \sum_{j=1,2}\bar A_j(P)\bigg[
{\cal B}_{ij}(Q) + \frac{1}{2}\frac{\partial  A_j(Q)}{\partial
Q_i}\bigg]+ \cdots.
\end{eqnarray}
This result can be used to write down  the required Hamiltonian system
to the second order in terms of  $A$'s. This is
 \begin{eqnarray}
{\cal H}_A  &=&  {\cal H} - \sum_i \left({\bar A}_i  {\bar u}_i - A_i  u_i \right)
+\frac{1}{2}\sum_{ij}\left[ {\bar A}_i {\bar A}_j \frac{\partial{\bar u}_i}
{\partial Q_j} +  A_iA_j\frac{\partial u_i} {\partial P_j} -2  {\bar A}_iA_j\frac{\partial u_j} {\partial Q_i} \right]
\nonumber\\
& & + \sum_{ij}A_j\left[
{\cal E}_{ij} - \frac{1}{2}\frac{\partial \bar A_j}{\partial
P_i}\right]\bar u_i - \sum_{ij}\bar A_j\left[
{\cal B}_{ij} + \frac{1}{2}\frac{\partial  A_j}{\partial
Q_i}\right] u_i
+{}\cdots
 \end{eqnarray}
where we the quantities $u_i$ and $\bar u_i$ are defined by
 \begin{eqnarray}
 u_i = \frac{\partial {\cal H}}{\partial P_i}, \qquad \bar u_i = \frac{\partial {\cal H}}{\partial Q_i}.
\end{eqnarray}
Here again bar is just a notation.
It is clear that the dressing transformation eliminates the fluctuations of the symplectic form,
 which become incorporated in the Hamiltonian.

\section{\bf Constant symplectic fluctuation}

\subsection{ Poisson structure}

As mentioned above the dressing transformation in the special case of a constant symplectic fluctuation can be
achieved by making use of the Hilbert--Schmidt procedure. This can be
seen as an exact alternative to one described in the former section.
From (65), one can verify that the matrix element
of the fluctuating tensors are
\begin{eqnarray}
{\cal E}_{ij} = \theta {\epsilon}_{ij}, \qquad {\cal B}_{ij} =
\bar{\theta}{\epsilon}_{ij}.
\end{eqnarray}
The nondegeneracy of $\omega$ is provided by the condition  $1 + \theta \bar
\theta \neq 0$. In addition, hereafter we assume that  $1 + \theta \bar
\theta  > 0$ is fulfilled. With the above particular modification of
the symplectic structure, the Poisson brackets (24-26) simply read as
\begin{eqnarray}
&& \{ q^i , q^j\} = - \frac{\theta}{1+\theta\bar{\theta}}\epsilon_{ij}
\\
&& \{ p^i , p^j\} = \frac{\bar{\theta}}{1+\theta\bar{\theta
}}\epsilon_{ij}
\\
&& \{ q^i , p^j\} = \frac{1}{1+\theta\bar{\theta }}\delta_{ij}
\end{eqnarray}
reflecting a deviation from the canonical brackets.

In this section, we  specify the form of the classical Hamiltonian.
More precisely,
we consider a bidimensional harmonic oscillator Hamiltonian of the type
\begin{eqnarray}
{\cal V}(p,q) = \frac{1}{2} \sum_{i} \left(p_i^2 + q_i^2\right).
\end{eqnarray}
This will be studied in subsection (4.3).

\subsection{ Dressing transformation and Quantization}

We start by noting that
 under the transformation
\begin{eqnarray}
&& Q^i = a q^i + \frac{1}{2} b\theta\sum_k\epsilon_{ki} p^k
\\
&& P^i = c p^i + \frac{1}{2} d\bar{\theta}\sum_k\epsilon_{ki} q^k
\end{eqnarray}
the Poisson brackets (73-75) give the canonical ones
\begin{eqnarray}
&& \left\{ Q^i , Q^j\right\} =  0 \nonumber \\
 && \left\{ P^i , P^j\right\} = 0 \\
 && \left\{ Q^i , P^j\right\} = \delta_{ij}\nonumber
\end{eqnarray}
once the real scalars $a$, $b$, $c$ and $d$ satisfy the following  set of
constraints
\begin{eqnarray}
&&4a^2 - 4ab-\theta\bar{\theta}b^2 = 0\nonumber\\
&&
4c^2 - 4cd-\theta\bar{\theta}d^2 = 0\nonumber\\
&&
4ac + 2\theta\bar{\theta}(ad+bc)-\theta\bar{\theta}bd = 4(1 +
\theta\bar{\theta}).\nonumber
\end{eqnarray}
A simple solution of such set is
\begin{eqnarray}
a = c = \frac{1}{b} = \frac{1}{d}=
\frac{1}{\sqrt{2}}\sqrt{1+\sqrt{1+\theta\bar{\theta}}}.
\end{eqnarray}
On the other hand, in terms of the above new dynamical variables, $\omega$
can be written as
\begin{eqnarray}
\omega = \sum_{i} dQ^i \wedge dP^i.
\end{eqnarray}
Inverting the  transformation (77-78), we obtain
\begin{eqnarray}
&& q^i = \frac{a}{\sqrt{1+\theta{\bar\theta}}}\left[ Q^i +
\frac{\theta}{2a^2} \sum_k\epsilon_{ik} P^k\right]
\\
&& p^i = \frac{a}{\sqrt{1+\theta{\bar\theta}}}\left[P^i +
\frac{\bar{\theta}}{2a^2}\sum_k\epsilon_{ik} Q^k\right].
\end{eqnarray}
For small values of $\theta$ and $\bar \theta$, we can
see that (82) and (83) give
\begin{eqnarray}
&& Q^i = \left(1 + \frac{1}{8} \theta \bar\theta \right) q^i +
\frac{\theta}{2}\sum_k\epsilon_{ki}p^k
\\
&&P^i = \left(1 + \frac{1}{8} \theta \bar\theta \right) p^i +
\frac{\bar\theta}{2}\sum_k\epsilon_{ki}q^k
\end{eqnarray}
which are sensitively comparable to the expressions (66) and (67).

\subsection{ New induced dynamics}

The Hamiltonian ${\cal V}$ (76)becomes
\begin{eqnarray}
{\cal V} = \frac{a^2}{2\left(1+\theta{\bar\theta}\right)} \left[ \sum_{i} \left(1 +
\frac{\theta^2}{4a^4}\right) P^iP^i + \left(1 +
\frac{\bar{\theta}^2}{4a^4}\right)Q^iQ^i + \left(\frac{\theta}{a^2} -
\frac{\bar{\theta}}{a^2}\right) \sum_j\epsilon_{ij}Q^iP^j\right].
\end{eqnarray}
Evidently the ($\theta ,\bar{\theta}$)-dependent terms
in (86) arise from the deformation of the symplectic structure. It
follows that  the deformation of the symplectic structure can be
thought as a perturbation reflecting the action of some external
potential on the system. This feature is very similar to the Landau problem in quantum mechanics.
For the purpose of the next section, we shall convert the
Hamiltonian (86) in complex notation. This can be achieved by
introducing the variables
\begin{eqnarray}
Z^i = \sqrt{\frac{\Delta}{2}} \left(Q^i + i \frac{P^i}{\Delta} \right), \qquad
 \bar Z^{i} = \sqrt{\frac{\Delta}{2}} \left(Q^i - i
\frac{P^i}{\Delta} \right)
\end{eqnarray}
where the involved parameter is
\begin{eqnarray}
\Delta = \sqrt{\frac{4a^4 +\bar{\theta}^2 }{4a^4 + \theta^2}}.
\end{eqnarray}
They satisfy the usual Poisson relations
\begin{eqnarray}
&& \left\{Z^i , Z^j \right\} = 0 \nonumber\\
&& \left \{Z^i , \bar Z^j\right\} =-i\delta_{ij}\nonumber\\
&& \left\{\bar Z^i , \bar Z^j\right\} = 0.\nonumber
\end{eqnarray}
The Hamiltonian ${\cal V}$ can be written as the sum of two contributions, such as
\begin{eqnarray}
{\cal V}-{\cal V}_0 = \frac{1}{4a^2}\frac{1}{1+\theta\bar{\theta}}
\sqrt{\left(4a^4+\theta^2\right) \left(4a^4 +\bar{\theta}^2  \right)}
\left(Z^{1}\bar Z^1 + Z^{2}\bar Z^2 \right)
\end{eqnarray}
where ${\cal V}_0$ is given by
\begin{eqnarray}
{\cal V}_0 = -\frac{i}{2}\frac{\theta - \bar \theta
}{1+\theta\bar{\theta}}\sum_{ij}
 \epsilon_{ij}\bar Z^{i}Z^j.
\end{eqnarray}
It can be also written in a form that is more appropriate
for our purpose. Indeed, by considering  new variables
\begin{eqnarray}
Z_+ = \frac{1}{\sqrt{2}} \left(Z^1 +i Z^2 \right), \qquad
Z_- = \frac{1}{\sqrt{2}} \left(Z^1 - i Z^2 \right)
\end{eqnarray}
and substituting (91) in (89-90), we end up with
\begin{eqnarray}
{\cal V} =  \left(\Omega - \delta \right) Z_+\bar Z_+ + \left(\Omega + \delta \right) Z_-\bar
Z_-
\end{eqnarray}
where $\Omega$ is
\begin{eqnarray}
\Omega = \frac{\sqrt{(4a^4+\theta^2)(4a^4 + \bar{\theta}^2
)}}{4a^2(1+\theta\bar \theta)}
\end{eqnarray}
and $\delta$ takes the form
\begin{eqnarray}
\delta = \frac{\theta - \bar \theta} {2\left(1+ \theta \bar \theta \right)}.
\end{eqnarray}
Note that, two-form (81) can be rewritten as
\begin{eqnarray}
\omega = i \left(dZ_+ \wedge d\bar Z_+ + dZ_- \wedge d\bar Z_- \right).
\end{eqnarray}
Upon quantization, all canonical variables become the Heisenberg operators satisfying commutation rules
 according to the canonical prescription, i.e. Poisson bracket $\longrightarrow$ -i commutator.
 It follows that the nonvanishing commutators  are
\begin{eqnarray}
\left[ Z_+ , \bar Z_+ \right] = 1, \qquad \left[ Z_- , \bar Z_-\right] = 1.
\end{eqnarray}
Note that, the Hamiltonian (92) is a superposition of two one dimensional harmonic oscillators.
Thus, the symplectic  modification induces a splitting of energy levels (degeneracy lifting).
This effect is very important and will have interesting consequences on the electromagnetic
excitations of quantum Hall effect in four-dimensional space. This is the main task
of the next section.

\section{Four-dimensional quantum  Hall droplet  }

\subsection{Brief review }

To illustrate the results of the previous sections, we consider a large number of particles,
evolving in four-dimensional complex projective
manifold ${\bf CP}^2$, under the action of a magnetic field generated by
two-form $\omega_0$ (12). In this situation the spectrum is highly degenerate,
splitting in Landau levels, and  it was shown [21] that there is
 one-to-one correspondence between the lowest Landau levels (LLL) or  ground state wavefunctions and
 the coherent states  given by (9),
  with $ d=2$  (${\cal F} \equiv {\rm LLL}$).
  For a strong magnetic field ($ k \to \infty $), the gap between Landau levels
becomes large and the particles are constrained to be accommodated in the LLL forming a quantum Hall droplet.

The dynamics of the droplet is characterized as follows. Since the LLL are highly
degenerated, one can fill states with $M = M_1+M_2$
particles where $M_i$ stands for the particle number in a given  mode
$i$. The corresponding density operator is then
\begin{equation}
\rho_0 = \sum_{n_1,n_2} \vert \ n_1 , n_2\ \rangle \ \langle \ n_1 , n_2\ \vert.
\end{equation}
The fluctuations, preserving the number of states, are described by an unitary transformation
\begin{equation}
\rho_0  \longrightarrow \rho =  U \rho_0 U^{\dagger}
\end{equation}
and the equation of motion is the quantum Liouville equation
\begin{equation}
i \frac {\partial \rho}{\partial t} = [ V , \rho ]
\end{equation}
where $V$ is the confining potential ensuring the degeneracy lifting of the LLL, see [21-22, 24] for more details.
Furthermore, since the LLL wavefunctions coincide with $SU(3)$ coherent states
in the symmetric representation, this offers a simple way to perform the semiclassical analysis.
This can be done by associating
 to every operator $A$ a symbol, such as
\begin{equation}
{\cal A}(\bar z, z) = \langle z | A | z \rangle = \langle 0
|\Omega^{\dag} A  \Omega| 0 \rangle.
\end{equation}
An associative
star product of two functions ${\cal A}(\bar z, z)$ and ${\cal
B}(\bar z, z)$ is then defined by
\begin{equation}
{\cal A}(\bar z, z)\star {\cal B}(\bar z, z) = \langle z | AB | z
\rangle
\end{equation}
which rewrites,  for large $k$, as
\begin{equation}
{\cal A}(\bar z, z)\star {\cal B}(\bar z, z) = {\cal A}(\bar z, z)
{\cal B}(\bar z, z) -
g^{j \bar m}
\partial_{j}{\cal A}(\bar z, z)\partial_{\bar m}{\cal B}(\bar z,
z).
\end{equation}
Then, the symbol or function associated with the commutator of two
operators $A$ and $B$ is given by
\begin{equation}
\langle z |[ A , B] | z \rangle = - g^{j \bar m}
\{\partial_{j}{\cal A}(\bar z, z)\partial_{\bar m}{\cal B}(\bar z,
z) - \partial_{j}{\cal B}(\bar z, z)\partial_{\bar m}{\cal A}(\bar
z, z)\}
\end{equation}
which leads to the result
\begin{equation}
\langle z |[ A , B] | z \rangle =  i \{{\cal A}(\bar z,
z), {\cal B}(\bar z, z)\} \equiv \{{\cal A}(\bar z, z), {\cal
B}(\bar z, z)\}_{\star}
\end{equation}
where $\{ , \}$ stands for the Poisson bracket defined by (13) and the notation $\{ ,
\}_{\star}$ stands for Moyal brackets.

With the above semiclassical correspondence, we can give
the symbol of the density matrix (97) in the limit of large number
of states, i.e. large magnetic field,
and large number of fermions $M$ ($ M < {\rm dim}{\cal F}$). This is  [21]
\begin{equation}
\rho_0(\bar z, z) \simeq \exp(-k\bar z\cdot z)\sum_{n=0}^M \frac{(k\bar
z\cdot z)^n}{n!}\simeq \Theta (M - k\bar z.z).
\end{equation}
where $\Theta$ is the usual step
function. It corresponds to an abelian
droplet configuration with boundary defined by $k\bar z\cdot z = M$ and
its radius is proportional to $\sqrt{M}$.

The confining potential can be defined in terms of the Fock number operators
$N_i \vert n_1 , n_2 \rangle = n_i \vert n_1 , n_2 \rangle$, with $ i= 1, 2$. This is
\begin{equation}
V = N_1 + N_2.
\end{equation}
The  associated symbol  is given by
\begin{equation}
{\cal V}(\bar z, z ) = \langle z |V| z \rangle = k \frac{\bar z\cdot z}{1-\bar z\cdot z}.
\end{equation}
which is exactly the potential given by (76).

This brief review gives the necessary tools needed to examine the electromagnetic
excitations of a quantum Hall droplet
in four-dimensional manifold by using the results obtained in the previous sections.
We will mainly focus on the situation where the matrix ${\cal B}$ and ${\cal E}$ are constants.

\subsection{Electromagnetic excitations of quantum Hall droplets}

It is clear that we may think the Hilbert ${\cal F}$ as the quantization of the phase space ${\bf CP}^2$ where
the symplectic form $\omega_0$ is proportional to the Kahler form on ${\bf CP}^2$.
The modification of the symplectic structure of the phase space induces electromagnetic interactions
of the quantum Hall droplets. The symplectic dressing methods, discussed previously, give
a prescription to eliminate the gauge fluctuations by encoding their effects in the
expression of the Hamiltonian of the system. Hence, in the case of constants ${\cal B}$
and ${\cal E}$, as shown above, the symplectic two form is mapped, via the relations
(82-83), (87) and (91), to its canonical form (95) in terms
of the new variables $Z_+$ and $Z_-$. The Poisson brackets become the canonical ones. Also, it is easily seen
 that the confining potential (107) can be mapped as
\begin{equation}
{\cal V}(\bar Z, Z ) = \Omega_+ Z_+\bar Z_+ + \Omega_- Z_-\bar Z_-
\end{equation}
where $\Omega_{\pm} = \Omega \mp \delta$ and the density function is given by
\begin{equation}
\rho_0(\bar Z, Z ) = \Theta\left[ M - k\left(\Omega_+ Z_+\bar Z_+ + \Omega_- Z_-\bar Z_-\right)\right].
\end{equation}
These are the main ingredients to evaluate the effective action describing
the quantum Hall droplets interacting with an external magnetic field $F$.
This action is given by [34]
\begin{equation}
S = \int dt {\rm Tr} \left[ \rho_0 U^{\dag}\left(i\partial_t - V\right)U \right].
\end{equation}
For a strong magnetic
field or $k$ large, the quantities appearing in this action can be
evaluated as classical functions.

Along similar lines as in [34, 21,24],
we start by computing the kinetic term. In this order, we set $U= e^{+i\Phi}$
$(\Phi^{\dag} = \Phi)$ to get
\begin{equation}
 i \int dt {\rm Tr}\left(\rho_0
U^{\dag}\partial_tU\right) \simeq \frac {1}{2k} \int d\mu
\{\Phi,\rho_0\}\partial_t\Phi
\end{equation}
where the symbol $\{ , \}$ is the
Poisson bracket. This gives
\begin{equation}
\{\Phi , \rho_0\} = (\Omega_+{\cal L}_+\Phi + \Omega_-{\cal L}_-\Phi)  \frac{\partial\rho_0}{\partial r^2}
\end{equation}
where $r^2 = \Omega_+ Z_+\bar Z_+ + \Omega_- Z_-\bar Z_-$ and
 the first order differential operators are defined by
\begin{equation}
 {\cal L}_{\alpha} =  i \left(Z_{\alpha} \frac{\partial}{\partial Z_{\alpha}}
- \bar Z_{\alpha}\frac{\partial}{\partial \bar Z_{\alpha}}\right), \qquad \alpha = + , -.
\end{equation}
In (112), the
 derivative of the density function  gives a  $\delta$ function with support on
the boundary $\partial {\cal D}$ of the droplet ${\cal D}$ defined
by $k r^2 = M$. Then, we have
\begin{equation}
 i \int dt {\rm Tr} \left(\rho_0
U^{\dag}\partial_tU \right) \approx -\frac{1}{2} \int_{\partial {\cal
D}\times{\bf R}^+} dt \left(\Omega_+{\cal L}_+\Phi + \Omega_-{\cal L}_-\Phi \right) \partial_t\Phi.
\end{equation}

We come now to the evaluation of the potential term in (110), which can be written as
\begin{equation}
Tr(\rho_0 U^{\dag} V U) = {\rm Tr} \left(\rho_0  V \right) + i {\rm Tr} \left(\left[\rho_0, V\right] \Phi\right)
+ \frac{1}{2} {\rm Tr} \left(\left[\rho_0, \Phi \right] \left[V, \Phi \right]\right) + \cdots.
\end{equation}
It can be easily verified that the first term in the second line in (115) gives a bulk contribution
 that can be ignored since we are interested to the edge dynamics. Further,
remark that it is $\Phi$-independent and contains no information
about the dynamics of the edge excitations.  From  (97) and (106), we have $[ \rho_0 , V ] = 0$, thus the second term in (115) vanishes.
The last term in (115) is evaluated  similarly to (114). Finally, we have
\begin{equation}
  \int dt {\rm Tr} \left(\rho_0
U^{\dag}{\cal H}U\right) \approx \frac{1}{2} \int_{\partial {\cal
D}\times{\bf R}^+} dt \left(\Omega_+{\cal L}_+\Phi + \Omega_-{\cal L}_-\Phi\right)^2.
\end{equation}

Combining (114) and (116), we get
\begin{equation}
S \approx -\frac{1}{2}\int_{\partial {\cal D}\times {\bf R }^+} dt
\left[{\Omega}_+ \left({\cal L}_+\Phi \right)
+ {\Omega}_-\left({\cal L}_-\Phi \right)\right]\left[ \left( \partial_t\Phi \right)
+{\Omega}_+ \left({\cal L}_+\Phi \right)
+ {\Omega}_-\left({\cal L}_-\Phi \right)\right].
\end{equation}
This action involves only the time derivative of $\Phi$ and the
tangential derivatives ${\cal L}_{\alpha}\Phi$. It is a generalization
of a chiral abelian Wess--Zumino--Witten (WZW) theory. For $\theta =
0 $ and $\bar\theta = 0$, we
recover the WZW usual action for the edge states associated with
un-gauged Hall droplets in four-dimensional space [21]. This is given by
\begin{equation}
S \approx -\frac{1}{2}\int_{\partial {\cal D}\times {\bf R }^+} dt
\left[(\partial_t\Phi ) ({\cal L}\Phi) +\omega ({\cal L}\Phi)^2
\right]).
\end{equation}
where ${\cal L} = {\cal L}_+ + {\cal L}_-$.

\subsection{ Edge fields}

The action (117) is minimized by the fields $\Phi$, which
are  satisfying
the equation of motion
\begin{equation}
\sum_{\alpha = \pm}(\Omega_{\alpha}{\cal L}_{\alpha})[\partial_t \Phi +\Omega_{\alpha}{\cal L}_{\alpha} \Phi] = 0.
\end{equation}
The edge field $\Phi$ can be expanded in powers of the phase space variables $Z_{\alpha}$.
Note that, since the excitations are moving on the real 3-sphere ${\bf S}^3 \sim SU(2)$,
it is convenient to introduce the $SU(2)$ parametrization. This is
\begin{equation}
\Omega_+ Z_+ = \sqrt{\frac{M}{k}}\frac{\sqrt{\bar \zeta \zeta}}{\sqrt{1+\bar \zeta \zeta}}e^{i\phi_+},
\qquad \Omega_- Z_- = \sqrt{\frac{M}{k}}\frac{1}{\sqrt{1+\bar \zeta \zeta}}e^{i\phi_-}
\end{equation}
where $\zeta$ and $\bar \zeta$ are the local complex coordinates for $SU(2)$.
The operators ${\cal L}_{\pm}$ reduce to partial
derivatives $\partial_{\phi_{\pm}}$ with respect to $\phi_{\pm}$. Thus, the field $\Phi$ is given as
\begin{equation}
\Phi = \sum_{n_+,n_-}  c_{n_+,n_-}(t) e^{i\phi_+n_+}e^{i\phi_-n_-}
\end{equation}
where the coefficients $c_{n_+,n_-}$ are $(\phi_+, \phi_-)$-independents for $(n_+ \neq 0, n_-\neq 0)$.
It follows that the
solution of the equation of motion (119) takes the form
\begin{equation}
\Phi = (\phi_+ - \Omega_+ t)+(\phi_- - \Omega_- t)+ \sum_{n_+ n_-}  c_{n_+,n_-}(0) e^{i(\phi_+ - \Omega_+ t)n_+}e^{i(\phi_- - \Omega_- t)n_-}.
\end{equation}
It is clear, from the last equation, that the noncommutativity arising from
the symplectic modification changes the propagation velocities of the edge field along the angular directions.
It is also important to stress that the velocities $\Omega_+$ and $\Omega_-$ are different
(respectively equal) for $\theta \neq \bar\theta$ (respectively $\theta = \bar\theta$).

\section{Concluding remarks}

We close the present analysis by summarizing the main points and results.
We first introduced the Bargman phase space of a quantum system whose elementary
excitations close the $su(3)$ Lie algebra. This space is interesting in three respects.
First, it equipped with a symplectic structure that one can vary in order to describe
the electromagnetic excitations of the system.
Second, the points of this space are in correspondence with the $SU(3)$ coherent states,
 which respect the over completion property. This provides us with an elegant tool to perform
 the semiclassical analysis (definition of star product and Moyal brackets).
Third, this phase space is four-dimensional manifold and one can consider a
 symplectic modification (17) such both positions $q$ and momentum $p$
 cease to Poisson commute. This can not be realized in two dimensional case.

In connection with this phase space, the present work addresses three major issues:
First,  the variation (or perturbation) of the symplectic two-form $\omega_0 \longrightarrow \omega_0 + F$,
which induces the noncommutative structures, can be eliminated through the Moser's lemma
 that  is a refined version of Darboux theorem. This leads to a dressing transformation (51), see also (68-69),
which converts the modified two-form in its undeformed form. The effects of the fluctuations become encoded
in the Hamiltonian of the system (70). The dynamics remains unchanged. We showed the dressing
transformation is equivalent to the Seiberg--Witten map (57-58).
This means that a symplectic modification and a noncommutative abelian gauge transformation are equivalents.

The second issue concerns the particular case where the matrix elements of the components
${\cal E}$ and ${\cal B}$ of electromagnetic fluctuation $F $ are constants (72).
We used the Hilbert--Schmidt orthonormalization
procedure to write down an exact dressing transformation (82-83).
Here again the effect of the non commutativity becomes encoded in the Hamiltonian (86).
This induces  the anisotropy of the harmonic oscillator potential (92)
and upon quantization generates a degeneracy lifting analogously to the well known Zeeman effect.

Finally, as application of the tools developed in this paper, we considered the problem of quantum Hall
effect in the complex projective space ${\bf CP}^2 = SU(3)/U(2)$. We derived the Wess--Zumino--Witten
action (117) governing the electromagnetic excitations of a large collection of fermions in
the lowest Landau levels. We obtained explicitly the edge field excitations (122)
traveling with modified velocities as consequence of the noncommutativity effects.

 \section*{ Acknowledgments}

MD would like to thank Max Planck Institute for Physics of
Complex Systems (Dresden-Germany) for  hospitality and support.
 He is also grateful to Abdus Salam International
 Centre for Theoretical Physics (Trieste-Italy).
AJ is grateful to
 Dr. Abdullah Aljaafari for his help and support.


\begin{thebibliography}{99}


\bibitem{1} N. Seiberg and E. Witten,  {\em JHEP} {\bf 9909} (1999) 032, [{{\sf
hep-th/9908142}}].

\bibitem{2} S. Doplicher, K. Fredenhagen and  J.E. Roberts,
{\em Commun. Math. Phys.}
  {\bf 172 } (1995) 187, [{{\sf hep-th/0303037}}].

\bibitem{3} M.R. Douglas and N.A. Neskrasov,  {\em Rev. Mod. Phys}. {\bf 73} (2001) 977, [{{\sf
hep-th/0106048}}].

\bibitem{4} R.J. Szabo,  {\em Phys. Rep}.  {\bf 378} (2003) 207,
[{{\sf hep-th/0109162}}].



\bibitem{5} V.P. Nair,   {\em Phys. Lett.} {\bf B505} (2001) 249,
  [{{\sf hep-th/0008027}}].

\bibitem{6} V.P. Nair and A.P. Polychronakos,  {\em Phys. Lett.} {\bf B505} (2001)
  267, [{{\sf hep-th/0011172}}]; A. Jellal, {\em  J. Phys. A: Math. Gen } {\bf 34} (2001) 10159,
[{{\sf hep-th/0502040}}].

\bibitem{7} O.F. Dayi and  A. Jellal, {\em Phys. Lett.} {\bf A287} (2001) 349, [{{\sf cond-mat/0103562}}];
{\em J. Math. Phys.} {\bf 43} (2002) 4592; (Erratum-ibid. {\bf 45} (2004) 827),  [{{\sf hep-th/0111267}}];
O.F. Dayi and L.T. Kelleyane, {\em Mod. Phys. Lett.} {\bf A17} (2002) 1937, [{{\sf hep-th/0202062}}].


\bibitem{8}   C. Duval and P.A. Horv\`athy,  {\em Phys. Lett.} {\bf B479} (2000 ) 284, [{{\sf hep-th/0002233}}];   {\em
  J. Phys. A: Math. Gen.} {\bf 34} (2001) 10097, [{{\sf hep-th/0106089}}];    {\em
   Phys. Lett.} {\bf B547 } (2002) 306, (Erratum-ibid. {\bf 588} (2004) 228), [{{\sf hep-th/0209166}}];
      P.A. Horv\`athy,   {\em  Ann. Phys.} {\bf 299} (2002) 128, [{{\sf hep-th/0201007}}];
  {\em
  SIGMA } {\bf 2}  (2006) 090, [{{\sf cond-mat/0609571}}];
{\em
   Phys. Lett.} {\bf A359}  (2006) 705,  [{{\sf cond-mat/0606472}}];
 P.A. Horv\`athy and  M.S. Plyushchay,  {\em  JHEP } {\bf 0206 } (2002) 033, [{{\sf hep-th/0201228}}];    {\em
 Nucl. Phys.} {\bf B714} (2005) 269, [{{\sf hep-th/0502040}}];
{\em
   Phys. Lett.} {\bf B595}  (2004) 547, [{{\sf hep-th/0404137}}].


  \bibitem{9} F. Delduc, Q. Duret, F. Gieres and M. Lefrancois,
  {\em J. Phys. Conf. Ser.} {\bf 103} (2008) 012020,  [{{\sf arXiv:0710.2239}}].


\bibitem{12} J. Moser,  {\em
  Trans. Amer. Math. Soc} {\bf 120} (1965) 286.

\bibitem{13} R.B. Laughlin,  {\em Phys. Rev.} {\bf B23}, (1981) 5632;  {\em Phys. Rev. Lett.}
{\bf 50} (1983) 1395.


\bibitem{14} L. Susskind, {\it The Quantum Hall Fluid and NonCommutative Chern--Simons Theory},
[{{\sf hep-th/0101029}}].

\bibitem{15}  A.P. Polychronakos,
 {\em
JHEP} {\bf 0104} (2001) 011, [{{\sf hep-th/0103013}}].

\bibitem{16}  S.C. Zhang and J.P Hu,
 {\em Science } {\bf 294} (2001) 823, [{{\sf cond-mat/0110572}}].

\bibitem{1} Y-X. Chen, B-Y. Hou and B-Y. Hou,  {\em Nucl. Phys.} {\bf B638} (2002) 220, [{{\sf
hep-th/0203095}}].

\bibitem{1}  M. Fabinger,  {\em   JHEP} {\bf 0205} (2002) 037, [{{\sf
hep-th/0201016}}].

\bibitem{1}H. Elvang and J. Polchinski, {\it The Quantum Hall Effect on ${\mathbb R}^4$}, [{{\sf
hep-th/0209104}}].

\bibitem{1} B.A. Bernevig, J.P. Hu, N. Toumbas and S.C. Zhang,  {\em Phys. Rev. Lett.} {\bf 91 } (2003) 236803, [{{\sf
cond-mat/0306045}}].

\bibitem{1} Brian P. Dolan ,  {\em  JHEP} {\bf 0305}  (2003) 018, [{{\sf
hep-th/0304037}}].

\bibitem{1} G. Meng,  {\em J. Phys. A: Math. Gen.} {\bf 36} (2003) 9415, [{{\sf
cond-mat/0306351}}].

\bibitem{17}   D. Karabali and V.P. Nair,
 {\em
Nucl. Phys.} {\bf B641} (2002) 533, [{{\sf hep-th/0203264}}];  {\em  Nucl. Phys.} {\bf B679} (2004) 427 , [{{\sf
hep-th/0307281}}];  {\em  Nucl. Phys. }
{\bf B697  } (2004) 513, [{{\sf hep-th/0403111}}];
 {\em J. Phys. A: Math. Gen.} {\bf  39} (2006) 12735, [{{\tt
hep-th/0606161}}].

\bibitem{1} V.P. Nair and S. Randjbar-Daemi,  {\em  Nucl. Phys. } {\bf B679} (2004) 447, [{{\sf
hep-th/0309212}}].


\bibitem{19}  A.P. Polychronakos,
 {\em  Nucl. Phys.} {\bf  B711} (2005) 505, [{{\sf
hep-th/0411065}}];  {\em  Nucl. Phys.} {\bf B705 } (2005) 457, [{{\sf
hep-th/0408194}}].

\bibitem{18} M. Daoud and A. Jellal,
 {\em Nucl. Phys. } {\bf
B764} (2007) 109, [{{\sf hep-th/0605289}}];  {\em Inter. J. Geom. Meth. Mod. Phys} {\bf 4} (2007)
1187, [{{\sf hep-th/0605290}}];  {\em Int. J. Mod. Phys.} {\bf A23} (2008)
3129, [{{\sf hep-th/0610157}}].


\bibitem{17}   D. Karabali,
 {\em  Nucl. Phys.} {\bf B726} (2005) 407, [{{\sf
hep-th/0507027}}];
 {\em Nucl. Phys.} {\bf B750} (2006) 265, [{{\sf hep-th/0605006}}].

\bibitem{1} B. Jurco, P. Schupp and J. Wess, {\em  Nucl. Phys} {\bf B584} (2000) 784, [{{\sf
hep-th/0005005}}].

\bibitem{1} B. Jurco, L. M\" oller, S. Schraml, P. Schupp and J. Wess,  {\em  Eur. Phys. J.} {\bf C21} (2001) 383,
 [{{\sf
hep-th/0104153}}].

\bibitem{1} W. Behr and A. Sykora,  {\em Nucl. Phys. } {\bf B698 } (2004) 473, [{{\sf
hep-th/0309145}}].

\bibitem{25} N. Jacobson, {\it Amer. J. Math.} {\bf 71} (1949) 149.

\bibitem{14}  T.D. Palev, {\it Lie Algebraical Aspects of the Quantum Statistics. Unitary Quantization (A-quantization)},  [{{\sf
hep-th/9705032}}].

\bibitem{15}   T.D. Palev and J. Van der Jeugt,  {\em   J. Math. Phys.} {\bf 43} (2002) 3850,  [{{\sf
hep-th/0010107}}].

\bibitem{24}    N.I. Stoilova and J. Van der Jeugt,  {\em   J. Math. Phys.} {\bf 46} (2005) 033501,  [{{\sf
math-ph/0409002}}];  {\em   J. Math. Phys.} {\bf 48} (2007) 043504,   [{{\sf
math-ph/0611085}}].

\bibitem{26}    M. Daoud, {\em   J. Phys. A: Math. Gen.} {\bf  39} (2006) 889,  [{{\sf
math-ph/0606050}}].

\bibitem{1} B. Sakita,  {\em Phys. Lett.} {\bf B387} (1996) 118, [{{\sf
hep-th/9607047}}].









\end{thebibliography}
\end{document}